\documentclass[twocolumn,preprint]{aastex6}

\usepackage{natbib}
\usepackage{bm}
\usepackage{amsmath}
\usepackage{rotating}
\newcommand{\code}[1]{\texttt{#1}}

\shorttitle{Outflow Column Density Profiles}
\shortauthors{Cottle, Scannapieco, and Br\"uggen}

\begin{document}


\title{Column Density Profiles of Cold Clouds Driven by Galactic Outflows}


\author{J'Neil Cottle, Evan Scannapieco,}
\affil{School of Earth and Space Exploration, Arizona State University, P.O. Box 871404, AZ 85287-1404, USA}


\and

\author{Marcus Br\"uggen}
\affil{Universit\"at Hamburg, Hamburger Sternwarte, Gojenbergsweg 112, D-21029, Hamburg, Germany}




\begin{abstract}

Absorption line studies are essential to understanding the origin, nature, and impact of starburst-driven galactic outflows. Such studies have revealed a multiphase medium with a number of poorly-understood features leading to a need to study the ionization mechanism of this gas. To better interpret these observations, we make use of a suite of adaptive mesh refinement hydrodynamic simulations of cold, atomic clouds driven by hot, supersonic outflows, including the effect of radiative cooling, thermal conduction, and an ionizing background characteristic of a starbursting galaxy.
Using a new analysis tool, {\it \textsc{trident}}, we estimate the equilibrium column density distributions for ten different ions: \ion{H}{1}, \ion{Mg}{2}, \ion{C}{2}, \ion{C}{3}, \ion{C}{4}, \ion{Si}{3}, \ion{Si}{4}, \ion{N}{5}, \ion{O}{6}, and \ion{Ne}{8}. These are fit to model profiles with two parameters describing the maximum column density and coverage, and for each ion we provide a table of these fit parameters, along with average velocities and line widths. Our results are most sensitive to Mach number and conduction efficiency, with higher Mach numbers and more efficient conduction leading to more compact, high column density clouds. We use our results to interpret down-the-barrel observations of outflows and find that the adopted ionization equilibrium model overpredicts column densities of ions such as \ion{Si}{4} and does not adequately capture the observed trends for \ion{N}{5} and \ion{O}{6}, implying the presence of strong non equilibrium ionization effects.

\end{abstract}


\section{Introduction}

It has been quite some time since galaxies have been studied as though they were island universes, growing in isolation by accreting material from their surroundings. Instead, it is now clear that the interactions between galaxies and their surrounding media are much more complex, depending on a network of feedback processes that are powered by stars \citep[e.g.][]{Dekel1986,MacLow1999,Scannapieco2001a,Scannapieco2001b,Mori2002,Scannapieco2002,Springel2003,DallaVecchia2008,Murray2011,Hopkins2012,Creasey2013,Muratov2015} and active galactic nuclei \citep[e.g.][]{Scannapieco2004,Sijacki2007,Schaye2015,Kaviraj2017}. One of the most important of these processes is the exchange of energy and material caused by galactic outflows. These outflows are thought to be driven by star formation and supernovae, \citep[e.g.][]{Heckman1990, Martin1999,Shapley2003,Martin2005,Veilleux2005} and can have a significant impact on the evolution of the galaxy, star formation rate and metallicities \citep[e.g.][]{Tremonti2004,Oppenheimer2010,Dave2011,Lu2015,Agertz2015}. Observations provide direct information on the
multiphase nature of these outflows \citep[e.g.][]{Sturm2011, Arav2013, Meiring2013, Bolatto2013, Kacprzak2014} as well as the composition and dynamics of the circumgalactic medium (CGM) into which they propagate \citep{Keeney2013, Rubin2014, Arribas2014, Werk2014, Wiseman2017}. However, disentangling the phases of the outflows and understanding their interactions with the environment has proven to be a challenge.

From an analytic perspective, \citet{Chevalier1985} derived a solution for a wind being driven from a region of uniform mass and continuous energy input. This model wind accurately describes the hot regions of galactic outflows observed in X-ray emission \citep{McCarthy1987}. However, these types of observations can only be made for nearby galaxies. For more distant objects, observations are limited to absorption measurements of colder gas, seen either in down-the-barrel observations of the host galaxy's background stellar continuum \citep{Chisholm2017} or along sightlines through the CGM of starburst galaxies with the background continuum provided by distant quasars (QSOs), \citep{HeckmanBorthakur2017, Borthakur2013}. While the two observations can provide information of the ionization and extent of the CGM, both are significantly limited in their ability to understand the dynamics of the outflowing material.
One particular anomaly in observations is the detection of absorption from both high ionization energy ions such as \ion{O}{6} at 138 eV and lower energies like \ion{Mg}{2} at 15 eV, with a distinct lack of absorption from \ion{N}{5} \citep[e.g.][]{Chisholm2017}. With an ionization energy around 97 eV, \ion{N}{5} is to be expected that the conditions which produce both \ion{O}{6} and \ion{Mg}{2} can also produce \ion{N}{5}. This discrepancy has been investigated for non-starbursting galaxies \citep{Werk2016}, but a cause in relation to starbursting galaxies in particular is yet to be determined. The number of direct observational predictions has been limited, making it unclear to what extent detailed models of the observational properties of cold clouds driven by galactic outflows can explain these trends.

Specifically, numerical simulations have focused on the nature of the outflowing material and the physics dominating the interaction between winds and cooler clouds.
\citet{Klein1994} have shown results for hydrodynamical simulations in which clouds within these winds were found to be accelerated and elongated over timescales longer than the time required for the shock to cross the cloud - demonstrating the need for longer simulations to fully understand the cloud evolution. Since then numerical simulations have expanded to investigate cloud-wind interactions from many angles from studies, including thermal conduction and radiative cooling \citep[e.g.][]{Orlando2005, Schneider2017}, accounting for non-equilibrium chemistry effects \citep[e.g.][]{Kwak2011}, and incorporating magnetic fields \citep[e.g.][]{MacLow1994, Fragile2005, McCourt2015}. Even so, these simulations have not covered the full parameter space relevant to galactic outflows and have not yet been directly connected to likely observations.

There have been several attempts to derive absorption line properties from cosmological simulations that include outflows \citep[e.g.][]{Oppenheimer2006, Oppenheimer2009, Ford2013,Turner2016}. However, these studies did not look at cold cloud properties with isolated outflows. Simulations of isolated cloud-wind systems have focus on the hydrodynamic interactions with less attention to the possible connections to observations. In addition, attempts to explain possible models for the spectra and absorption profiles observed fall short of having reliable ways to connect to simulations already performed.

New analytic tools, such as \textsc{trident} \citep{Hummels2017}, can help bridge the gap between simulations and observations. This can be done by generating synthetic spectra and calculating ion number densities within simulations without the extra computational cost of including a chemistry solver. For this work, we explore the possibility of generating column densities of commonly observed ions from existing simulations with \textsc{trident} in order to make comparisons between simulation results and actual outflow observations.

In this paper we present synthetic column density calculations and velocity profiles of clouds simulated with both radiative cooling and thermal conduction at various evolutionary stages. In Section \ref{sec:sim} we discuss the simulations used within this study including the parameters and relevant physics. Within Section \ref{sec:method} we outline the methods of calculating the column density and velocity profiles as well as the procedure for fitting profiles for each cloud. Section \ref{sec:results} includes comparisons across simulation parameters and ion species, with an application of these results to observations in Section \ref{sec:app}. We conclude with a discussion and motivation for future work in Section \ref{sec:discuss}.

\section{Simulations}\label{sec:sim}
We performed a full analysis of the ion densities on the outflow simulations in \citet{Scannapieco2015} and \citet{Bruggen2016}, SB15 and BS16 respectively hereafter. This suite of simulations was carried out with FLASH (version 4.2) \citep{Fryxell2000}, a multidimensional hydrodynamics code that solves the fluid equations on a Cartesian grid with a directionally split Piecewise-Parabolic Method \citep{Colella1984}. The simulations were done in three dimensions, as limiting the degrees of freedom can influence the development of shear instabilities.
They assumed an initial cloud radius of 100 parsec, a temperature of 10$^4$ K, and a mass density of $\rho = 10^{-24}$ g cm$^{-3}$ and a mean atomic mass of $\mu = 0.6$. These parameters result in a total column density of $3.1\times 10^{21}$ cm$^{-2}$. As shown below, this column density determines cloud evolution, rather than the radius and density.

Initially, the cloud was positioned at (0, 0, 0) within the domain covering a physical volume of $-800 \times 800$ parsec in the $x$ and $y$ directions and $-400 \times 800$ parsec in the $z$ direction, which was the direction of the hot outflowing material. The interaction at the $z$ boundary was defined by a condition where the incoming material is added to the grid and given the same values of density, $v_{\rm hot}$, and $c_{\rm s,hot}$ as the initial conditions. For all other boundaries, the FLASH ``diode" condition was used, which assumes the gradient normal to the edge of the domain of all variables except pressure to be zero and only lets material flow out of the grid.

\subsection{Physics of Cloud Evolution}
Two important timescales that influence the evolution of a cold cloud embedded within a hot wind and the cloud crushing time, $t_{\rm cc},$ and the cooling time, $t_{\rm cool}$ . The cloud crushing time effectively describes the amount of time it would take the initial shock to travel halfway through the cloud and is given by
\begin{equation}
t_{\rm cc} = \frac{R_c}{v_{\rm hot} \chi_0^{1/2}},
\end{equation}
\\
which is dependent only on the velocity of the wind, $v_{\rm hot}$, and the density ratio, $\chi_0$ \citep[e.g.][]{Klein1994}. The cooling time, which determines the time for the cloud to radiate away its thermal energy is given by
\begin{equation}
t_{\rm cool} = \frac{(3/2)n_c k \textbf{T}}{\Lambda(T) n_{e,c} n_{i,c}},
\end{equation}
\\
where $T$ is the temperature and $\Lambda(T)$ is the equilibrium cooling function at $T$ with $n_{c}$, $n_{e,c}$ and $n_{i,c}$ are the total, electron and ion number densities within the cloud. If the ratio of $t_{\rm cool}/t_{\rm cc} = N_{\rm cool}/(n_{i,c} r_c)$ with $N_{\rm cool} \equiv 3 k T v n_v (2 \Lambda \chi^{1/2} n_{e,c})^{-1}$ is below one, then cooling will have a significant influence of the evolution of the cloud.
Table 1 in SB15 gives values for $N_{\rm cool}$ as calculated using equilibrium cooling cures from \citet{Wiersma2009} assuming solar metallicity and a mean molecular mass of 0.6. With column densities between 10$^{17}$ and 10$^{19}$ cm$^{-2}$, the resulting ratio between cooling time to cloud crushing time is small. For the range of parameters used, the clouds are able to cool on a timescale much shorter than the timescale for the evolution of the cloud allowing for cooling to influence the cloud before it is disrupted by the shock.

Within the simulations, cooling was computed in the optically thin limit assuming local thermodynamic equilibrium
\begin{equation}
\dot{E}_{\rm cool} = (1-Y) \left(1 - \frac{Y}{2} \right) \frac{\rho \Lambda}{(\mu m_p)^2},
\end{equation}
where $\dot{E}_{\rm cool}$ is the radiated energy per unit mass, $\rho$ is the density in the cell, $m_p$ is the proton mass, $Y = 0.24$ is the helium mass fraction, $\mu = 0.6$ the mean atomic mass, and $\Lambda(T, Z)$ is the cooling rate as a function of temperature and metallicity. With the assumption that the abundance ratios of the metals are always solar, the tables compiled by \citet{Wiersma2009} were used. Heating by a photoionizing background was not included in the calculations, however sub-cycling was implemented \citep{Gray2010} along with a cooling floor at $T=10^4$K.

The fluid equations including thermal conduction and radiative cooling as solved by FLASH are
\begin{equation}
\partial_t \rho + \nabla \cdot (\rho \bm{u}) = 0,
\end{equation}
\begin{equation}
\rho [\partial_t \bm{u} + (\bm{u} \cdot \nabla)\bm{u}] = -\nabla p,
\end{equation}
\begin{equation}
\partial_t E + \nabla \cdot [E\bm{u}] = - \nabla \cdot (p\bm{u}) - n^2 \Lambda(T) + \nabla \cdot \bm{q},
\end{equation}
with $\rho$ the density, $\bm{u}$ the velocity, $p = k_B T \rho/(\mu m_p)$ the pressure and $E = p/(\gamma - 1) + \frac{1}{2} \rho |\bm{u}|^2$ the total energy density, $\Lambda(T)$ is the radiative cooling function and $\bm{q}$ describes the heat flux due to conduction. We adopt a saturated thermal conduction limit when the mean free path of electrons is much larger than the length scale of the temperature gradient. This leads to the definition
\begin{equation}
\bm{q} = \text{min} ( \kappa(T) \nabla T, \quad 0.34 n_e k_B T c_{s, e} \nabla T | \nabla T | ),
\end{equation}
\citep{Cowie1977}, where $\kappa (T) = 5.6 \times 10^{-6} T^{5/2}$ erg s$^{-1}$ K$^{-1}$ cm$^{-1}$ and $c_{s,e} = (k_BT/m_e)^{1/2}$ is the isothermal sound speed of the electrons in the wind with $m_e$ the mass of the electron. It is assumed that electrons and ions have the same temperature. The diffusion equation describing conduction is then solved with the general implicit diffusion solver in FLASH. Saturated thermal conduction was also implemented with the use of a flux limiter that modifies the diffusion coefficient to vary until some maximum flux as determined by the Larsen flux limiter (Morel 2000).
In units of cloud crushing times, these equations are invariant under the transformation
\begin{equation}
\bf{x} \rightarrow \alpha \bf{x}, \qquad \bf{t} \rightarrow \alpha \bf{t}, \qquad \text{and} \qquad \rho \rightarrow \alpha \rho
\end{equation}
resulting in the evolution of the cloud only depending on the product of the size and density.

\subsection{Selection of Evolutionary Stages}
While the cloud crushing time is an good description of the disruption time for a single cloud, this study compares evolutionary stages across many types of clouds. To compare to $t_{\rm cc}$, another timescale is defined over the course of the cloud's evolution based on the mass fraction of the cloud that is at or above 1/3 of the cloud's original density. The first time, $t_{95},$ corresponds to the time at which 95\% of the cloud is at or above this density. The following three times, $t_{75}$, $t_{50}$ and $t_{25}$ follow a similar pattern with 75\%, 50\% and 25\% of the cloud. These four stages correspond to the four evolutionary stages we consider while estimating column densities.

\subsection{Frame Changing and Refinement/Derefinement}

In order to follow the disruption of the clouds, it was necessary for the simulations to shift frames as the cloud drifts through the wind. To do this SB15 and BS16 have implemented an automated frame change routine (see SB15 and BS16 for details). In addition, they used FLASH's default variables of temperature and density with a refinement criterion on 0.8. A secondary refinement condition was enforced to ensure the simulation maintained high resolution in areas important to cloud evolution and to reduce the computational cost of higher refinement in areas of the simulation that have less influence on the cloud evolution. This additional condition imposed derefinement on cells that satisfied one of the following (1) the cell was outside of a cylinder along the $z$ axis with radius three times the initial cloud raids or nine times the current $x$ extent of the cloud or (2) the cell was outside of a cylinder centered on the $z$ axis with radius equal to the initial cloud radius or three times the current $x$ extent of the cloud \textit{and} both the distance to the $x-y$ plane and the $z$ center of the cloud were greater than three times the current radius of the cloud.

\subsection{Parameters}

The parameter space for these simulations to be reduced to the wind parameters; $T_{\rm hot}$, $v_{\rm hot}$, and column density of the cloud. According to \citet{Chevalier1985}, Mach number depends only on $r/R_\star$ where $r$ is the distance from the outflowing region and $R_\star$ is the driving radius of the flow. This radius reflects the size of the region in which the energy input from sources such as supernovae accelerates the gas. At the edge of this region the gas becomes supersonic and tends to a constant velocity a further radii. For M82, $R_\star \approx 300$ pc. \citep{McKeith1995} It follows that the energy and mass input from the wind can be fully described with the velocity of the hot medium while the Mach number corresponds to sampling the wind as a function of radius. For the Mach numbers considered, assuming $R_\star = 300$ pc the physical scale of these radii range from 0.3 to 2.9 kpc from the central starburst. For the cloud with a temperature corresponding to the minimum temperature attainable with atomic cooling ($\approx 10^4$ K), the Jeans length for this gas is $\lambda_J \approx 2$ kpc, much larger than the size of the clouds considered, indicating that the clouds must be confined by pressure to keep from expanding. The pressure equilibrium then requires the ratio of the cloud density to the wind density, $\chi_0$, to be equal to the ratio of the temperature of the wind to the temperature of the cloud.

The choices for $T_{\rm hot}$ and $v_{\rm hot}$, as well as the corresponding density contrast and cloud crushing times, for a cloud radius of 100 parsec are given Table \ref{tab:params} for, both, the cooling and conduction runs. The Mach number of the hot wind, $M_{\rm hot}$ is also given. The naming scheme of the runs describes Mach number, wind velocity and wind temperature in order, with suffixes denoting other unique traits of the run. The parameters were chosen to focus on regions outside of the driving radius, $r > R_\star$, with Mach numbers $\geq 1$ and provide multiple runs with the same temperatures and velocities to study the impact of changing the Mach number within the hot wind. Also included are runs with both wind and cloud densities 10 times greater than their original counterparts (named with the -hc suffix) and one low conduction run with one third the Spitzer value used in all other conduction runs (named with a -lc suffix).

\begin{table*}
\caption{Simulation parameters - conduction runs end in -c; high column in -hc; low conduction in -lc \label{tab:params} }
\centering
\begin{tabular}{lccccccc}
Name & Conduction & $M_{\rm hot}$ & $v_{\rm hot}$ & $T_{\rm hot}$ & $ T_{\rm hot}$ & $\chi_0$ & $t_{\rm cc}$\\
& & & (km s$^{-1}$) & ($10^6$ K) & (keV) & & (Myr/100pc)\\
\hline\\
M0.5-v430-T3 & & 0.5 & 430 & 30 & 2.7 & 3000 & 12.5\\
M1-v480-T1 & & 1 & 480 & 10 & 0.86 & 1000 & 6.4\\
M1-v860-T3 & & 1 &860 & 30 & 3.7 & 3000 & 6.2\\
M1-v1500-T10 & & 1 & 1500 & 100 & 8.6 & 10000 &6.5\\
M3.8-v1000-T0.3 & & 3.8 & 1000 & 3 & 0.27 & 300 &1.7\\
M3.5-v1700-T1 & & 3.5 & 1700 & 10 & 0.86 & 1000 & 1.8\\
M3.6-v3000-T3 & & 3.6 & 3000 & 30 & 2.7 & 3000 & 1.8\\
M6.5-v1700-T0.3 & & 6.5 & 1700 & 3 & 0.27 & 300 & 1.0\\
M6.2-v3000-T1 & & 6.2 & 3000 & 10 & 0.86 & 1000 & 1.0\\
M11.4-v3000-T0.3 & & 11.4 & 3000 & 3 & 0.27 & 300 & 0.56\\
M1-v480-T1-c & \checkmark & 1 & 480 & 10 & 0.86 & 1000 & 6.4\\
M1-v860-T3-c & \checkmark & 1 &860 & 30 & 3.7 & 3000 & 6.2\\
M1-v1500-T10-c & \checkmark & 1 & 1500 & 100 & 8.6 & 10000 &6.5\\
M3.8-v1000-T0.3-c & \checkmark & 3.8 & 1000 & 3 & 0.27 & 300 &1.7\\
M3.5-v1700-T1-c & \checkmark & 3.5 & 1700 & 10 & 0.86 & 1000 & 1.8\\
M3.6-v3000-T3-c & \checkmark & 3.6 & 3000 & 30 & 2.7 & 3000 & 1.8\\
M6.5-v1700-T0.3-c & \checkmark & 6.5 & 1700 & 3 & 0.27 & 300 & 1.0\\
M11.4-v3000-T0.3-c & \checkmark & 11.4 & 3000 & 3 & 0.27 & 300 & 0.56\\
M3.8-v1000-T0.3-hc & \checkmark & 3.8 & 1000 & 3 & 0.27 & 300 & 1.7\\
M3.5-v1700-T1-hc & \checkmark & 3.5 & 1700 & 10 & 0.86 & 1000 & 1.8\\
M3.6-v3000-T3-hc & \checkmark & 3.6 & 3000 & 30 & 2.7 & 3000 & 1.8\\
M6.5-v1700-T0.3-lc & \checkmark & 6.5 & 1700 & 3 & 0.27 & 300 & 1.0\\
\end{tabular}
\end{table*}

\section{Estimation of Observables} \label{sec:method}
\subsection{Trident Analysis}
Our analysis makes use of the \textsc{trident} analysis tool \citep{Hummels2017}, an extension of the yt analysis code \citep{yt2011}. \textsc{trident} is a multifunctional tool created to enable simulated observations of astronomical hydrodynamic simulations. It can be used to create absorption line spectra through simulated datasets as well as column density maps for ion species not originally within the simulation outputs. The full description of the code can be found in \cite{Hummels2017}. However, a short description of the relevant details is given here.

In order to generate density maps and spectra, \textsc{trident} first calculates the density of a given ion within the simulated dataset. This is done through the module \code{ion balance}. The module first determines if the dataset contains a density element for each cell within the domain considered, this may be the entire dataset or a subset representing a sightline as defined by \textsc{trident}'s \code{LightRay}. If the simulation explicitly tracks the ion's density through a chemistry solver, this density is used. However, for this paper, each ion number density is derived from the gas density and metallicity fields within the dataset and an ionization fraction assuming ionization equilibrium. The final estimation for the number density of the \textit{i}-th ion of element $X$ becomes
\begin{equation}
n_{X,i} = \bm{f_H} \frac{\rho}{m_H}Z \left(\frac{n_X}{n_H} \right)_\odot f_{X,i},
\end{equation}
\\
where $\rho$ and $Z$ are the gas density and metallicity fields, respectively, from the dataset,
$\bm{f_H}$ is the primordial H mass fraction with an adopted value of 0.76, and $Z(\frac{n_X}{n_H})_\odot $ is the solar abundance.

\subsection{UV Background}

The equilibrium ionization fraction, $f_{X,i}$, is a function of temperature, density and incident radiation, most typically a UV metagalactic background. For use within the \code{ion balance} module the ionization fraction is linearly interpolated over a grid of pre-calculated ionization fractions through temperature, density and redshift. The default UV background for \textsc{trident} is the \citet{Haardt2012} metagalactic background. While this is appropriate for approximating the ions within the intergalactic medium, it is not an accurate estimation of the environment around starburst galaxies.

In order to create a new ionization fraction lookup table to integrate with \textsc{trident}, the shape and intensity of the incident radiation was based off of a STARTBURST99 model \citep{Leitherer1999}. Here we used the best-fit theoretical model found within \citet{Chisholm2017} from `down-the-barrel' observations of the outflow in galaxy J1226+2152. Such an orientation allowed for the absorption lines of the ISM to be embedded within the stellar continuum. The best-fit model was found by fitting both the continuum and extinction using a Calzetti extinction law \citep{Calzetti2000}. The STARBURST99 models make use of the Geneva stellar evolution model and varied interstellar continuum metallicities. The best-fit model had a stellar metallicity of 0.2 $Z_\odot$ and a light-weighted age of 11 Myr. With the shape of the incident radiation given by the best-fit STARBURST99 model, the strength of the radiation is dependent on the distance from the source, which we infer from the measured ionization parameter in \citet{Chisholm2017}, $\log(U) = -2$.

\begin{figure*}
\includegraphics[angle=0,width=0.95\textwidth]{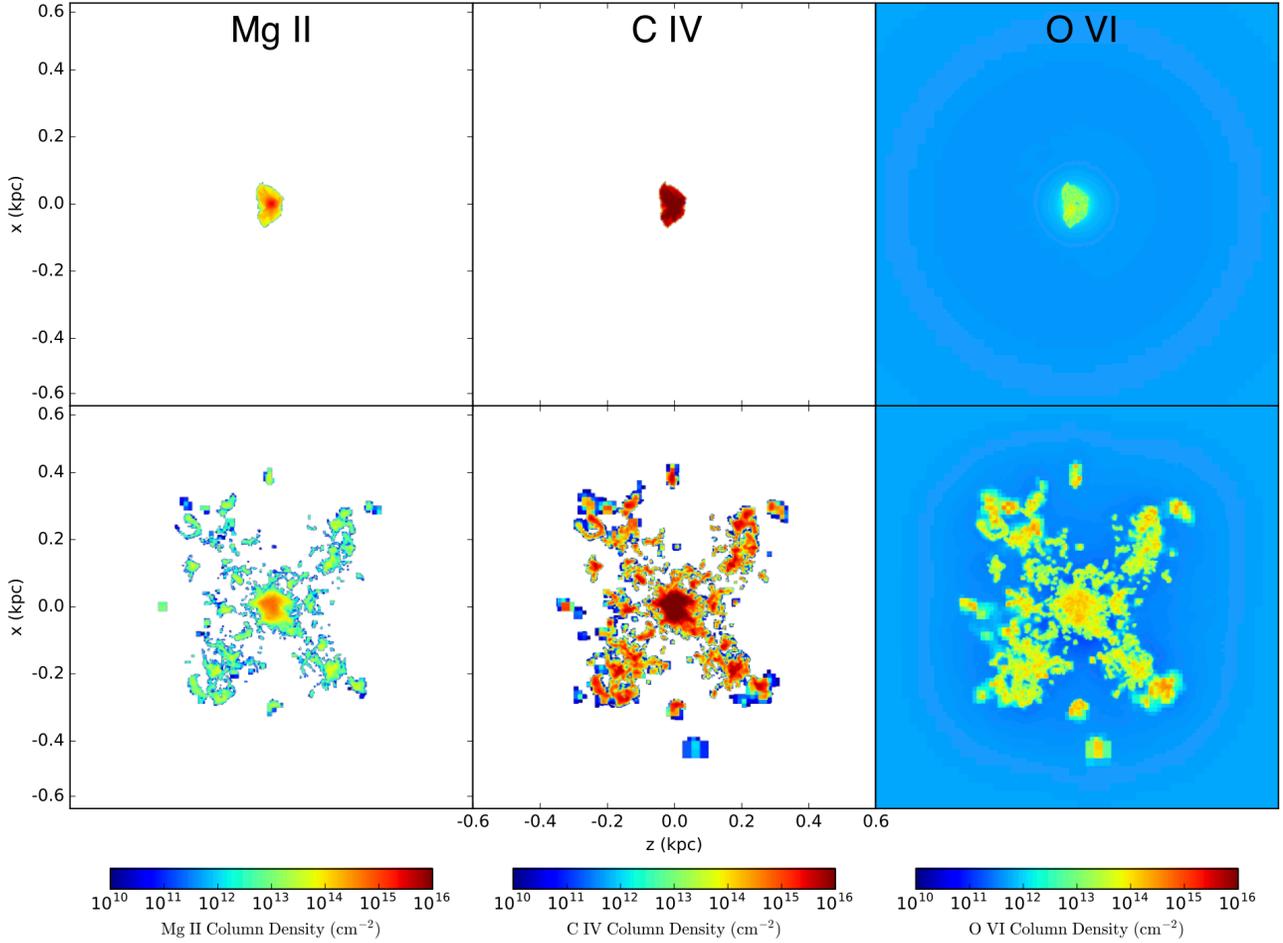}
\caption{Density projections down-the-barrel (through the wind), for \ion{Mg}{2} (left), \ion{C}{4} (middle), and \ion{O}{6}(right) at $t_{50}$ for the M3.8-v1000-T0.3 runs. The efficient conduction run is shown in the top panels and inefficient conduction is shown in the bottom panels. \label{fig:projections}}
\end{figure*}

The construction of the table then followed the same procedure as outlined in \citet{Smith2008, Smith2017}. Ionization fractions were computed within a grid containing temperature, hydrogen number density and redshift. While redshift was not explicitly taken into account, these results are applicable to observations with $z \geq 0$ due to the fact this analysis focused on the influence of the dominant starburst background which is redshift independent. The grid was populated with calculations using the photoionization software, CLOUDY \citep{Ferland2013}, which takes the best fit STARBURST99 model as the shape of incident radiation and an ionization parameter of $\log(U) = -2$ \citep[see ][]{Chisholm2017}. These simulations spanned a range of temperatures from 10 to 10$^9$ K, in step sizes of 0.025 dex, and hydrogen number densities 10$^{-9}$ to 10$^4$ cm$^{-3}$, in step sizes of 0.125 dex, to mimic the default table within \textsc{trident}, and allow for integration with the existing \textsc{trident} procedure with little modification. The newly generated table was then loaded in place of the default ionization table and ion number densities were calculated as described above.

With the use of the new table, column densities maps can be created as shown in Figure \ref{fig:projections}. These projections are for the M3.8-v1000-T0.3-c and M3.8-v1000-T0.3 runs highlighting the difference in structure between the conduction and cooling runs. While conduction creates dense clouds, runs without conduction are much more diffuse with more coverage. With the inclusion of conduction, higher column densities are possible than those that can result from shock-induced ionization. For these runs, compression by the evaporative flow is most significant at low Mach numbers and low density contrasts. However, as BS16 note, evaporative compression influences all runs producing small dense clouds at late times in parallel with the streamwise pressure gradient stretching the clouds into filaments. While the filaments are created in the runs with radiative cooling and inefficient conduction, it is this evaporation that leads to such different morphologies between the two sets of simulations. If conduction is suppressed by factors such as magnetic fields, the radiative cooling runs may lead to clouds more descriptive of the accelerated material in outflows. 

\begin{figure}
\includegraphics[angle=0,width=\columnwidth]{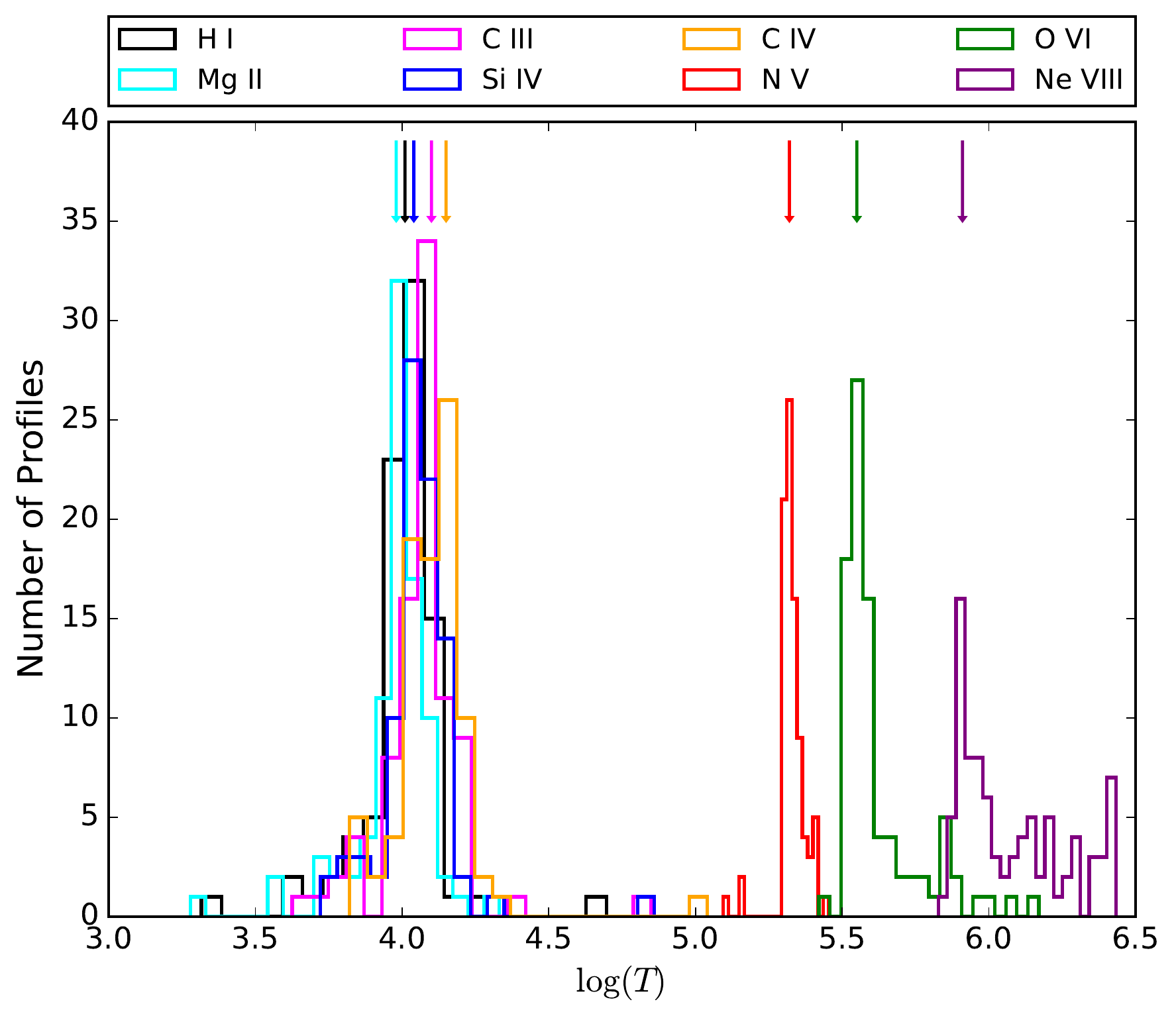}
\caption{The distribution of average temperatures for 8 ions; \ion{H}{1}, \ion{Mg}{2}, \ion{C}{3}, \ion{C}{4}, \ion{Si}{4}, \ion{N}{5}, \ion{O}{6} and \ion{Ne}{8} over all times and runs. An approximate average temperature for each ion is highlighted with an arrow above the distribution. \label{fig:temphist}}
\end{figure}

The multiphase nature of these outflows can be seen by looking at the average temperature for each ion. In Figure \ref{fig:temphist} the distribution of average temperature across all runs and times for ions of various ionization potentials is shown. Despite the wide range of wind parameters, there is a distinct distribution of temperatures for each ion. It is particularly interesting that the average temperature for \ion{O}{6} and \ion{N}{5} are substantially higher than the temperature of the cloud, indicating that these ions are produced in a separate phase than the low and intermediate ionization potential ions such as \ion{Mg}{2} and \ion{Si}{4}. These high temperatures imply that the higher ionization potential ions are primarily collisionally ionized within these simulations.

\subsection{Column Density Profiles}
The column density for each projection along the sightline for each ion at any stage was computed as
\begin{equation}
N = \int_{\rm los} n_i(z) dz,
\end{equation}
where $n_i(z)$ is the number density of ion, $i$, and $z$ is the direction of the projection.

\begin{figure}
\includegraphics[angle=0,width=\columnwidth]{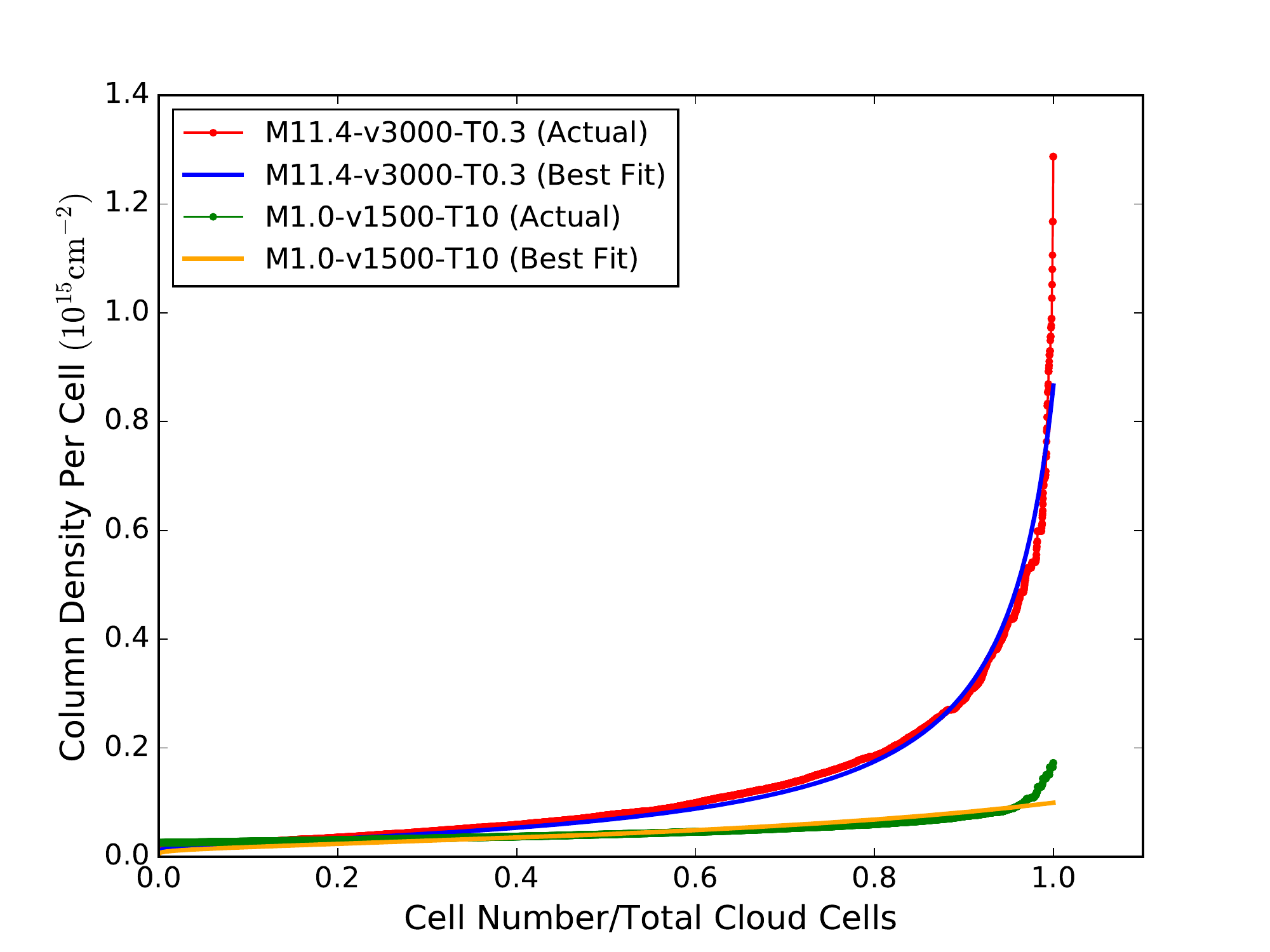}
\caption{Sample of the column density profile with the best-fit line over plotted. Two runs are shown at $t_{75}$ for the ranked column density of OVI; M11.4-v3000-T0.3 (actual in red, best-fit in blue) and M1.0-v1500-T10 (actual in green, best-fit in yellow). The best-fit overplotted is the functional form of equation (\ref{eq:tauFunction}) with the best-fit parameters $N_0 = 8.68 \times 10^{14}$ cm$^{-2}$ and $q=0.18$ for the M11.4-v3000-T0.3 run and $N_0 = 9.94 \times 10^{13}$ cm$^{-2}$ and $q=0.02$ for the M1.0-v1500-T10 run.  \label{fig:opticalDepthRank}}
\end{figure}

We considered a down-the-barrel projection representative of outflows for all runs.
Each projection was taken at a fixed resolution of 800 $\times$ 800 cells for the domain covering 1.6 kpc $\times$ 1.6 kpc, resulting in one pixel per 4 pc$^2$. At this resolution the initial cloud area within the simulation covered 0.0314 kpc$^2$, or 7880 cells. For each cell we ranked the column densities from lowest to highest and took the top 7880 to give a profile of the column densities that cover a simulation area equal to the initial cloud area. An example of this profile can be seen in Figure \ref{fig:opticalDepthRank}. We determine the profiles can be described by the functional form
\begin{equation}\label{eq:tauFunction}
N(x) = N_0 \frac{0.01}{1.01-x^q},
\end{equation}
where $x$ is the fractional rank of each cell expressed as a fraction of the total pixels and $N_0$ translates to a upper limit on column density for the cloud and $q$ expresses the degree to which the cloud has been compacted. A high $q$ relates to a very compact cloud along the line of sight, while a low $q$ is more descriptive of a diffuse cloud along the line of sight or a consistent column density throughout the entire simulation area considered. This parameter can also be thought of as an analog for coverage, with high $q$ translating to a small amount of coverage with nearly maximum column density and low $q$ describing greater coverage of the sightline containing high column density.

We found the posterior probability distributions of the parameters $N_0$ and $q$ with an Affine Invariant Markov Chain Monte Carlo Ensemble sampler through the use of the Python package \code{emcee} developed by \citet{Foreman2013}. With the use of priors, $N_0$ and $q$ were restricted to values between 0 and twice the maximum column density in the domain and 0 and 100, respectively. The 1$\sigma$ errors were then derived from the 16 and 84th percentile contours of the posterior.

It is also informative to consider a total average column density. However, as the column density profiles have been cast into a space relating to a fraction of the initial cloud area ($x$), this average is restricted to considering this specified simulation area rather than the entire cloud. An average column density over a simulation area equal to the size of the initial cloud area was therefore calculated as a proxy for total average column density. This was determined numerically from the best-fit model profile as
\begin{equation}
N_{\rm total} = \int_0^1 N_0 \frac{0.01}{1.01-x^q} dx.
\end{equation}
The column density of a portion of the cloud can be found in a similar way, changing the limits of integration to reflect the portion of the cloud considered. The column density of the densest half of the cloud corresponds to the value of the integral above with limits from 0.5 to 1.
The average column density, along with the best-fit parameters and the corresponding errors, was stored in a lookup table.

Also calculated was the average velocity of each ion as well as an estimate of the width of an approximate Gaussian profile, or $b$ parameter, including both the thermal velocity spread ($b_t$) and contribution of Doppler broadening ($b_d$). The average velocity, weighted by the ion number density was calculated as
\begin{equation}
\bar{v_i} = \frac{\int n_i(x, y, z) v(x) dV}{\int n_i(x, y, z) dV}.
\end{equation}
Again, with $x$ the direction along the line-of-sight. The $b$ parameter was estimated to be the average of the $b$ calculated for all sightlines for the given projection, expressed as
$
b = \left<\sqrt{b_t^2 + b_d^2}\right>,
$
with
\begin{equation}
b_t^2 = \frac{\int \frac{2k_b T(x, y, z)}{\mu_i m_p} n_i(x, y, z) dx}{\int n_i(x, y, z) dx}
\end{equation}
and
\begin{equation}
b_d^2 = \frac{\int (v(x) - \bar{v_i})^2 n_i(x, y, z) dx}{\int n_i(x, y, z) dx}.
\end{equation}
Here $T(x,y,z)$ is the temperature of the gas and $\mu_i$, the mass number of the ion and $m_p$, the mass of a proton, with all other constants defined in the usual form. The values for $\bar{v_i}$ and $b$ are stored within the table of best-fit parameters.

This table is created for 10 different ions including low ionization energies prevalent within the cool cloud material such as \ion{H}{1} and \ion{Mg}{2} through intermediate energies, \ion{Si}{3}, \ion{Si}{4}, \ion{C}{2}, \ion{C}{3}, \ion{C}{4}, and \ion{N}{5} to those at the high end, \ion{O}{6} and \ion{Ne}{8}.
These best-fit parameters and the associated errors are quoted in 10 digital tables, one for each ion considered. Table \ref{tab:paramSample} represents a sample of these tables. The full tables are available online \footnote[1]{$\rm www.public.asu.edu/\sim jcottle1/coldensprofiles.html$}.

\section{Results} \label{sec:results}

\begin{figure*}
\centering
\includegraphics[angle=0,scale=0.6]{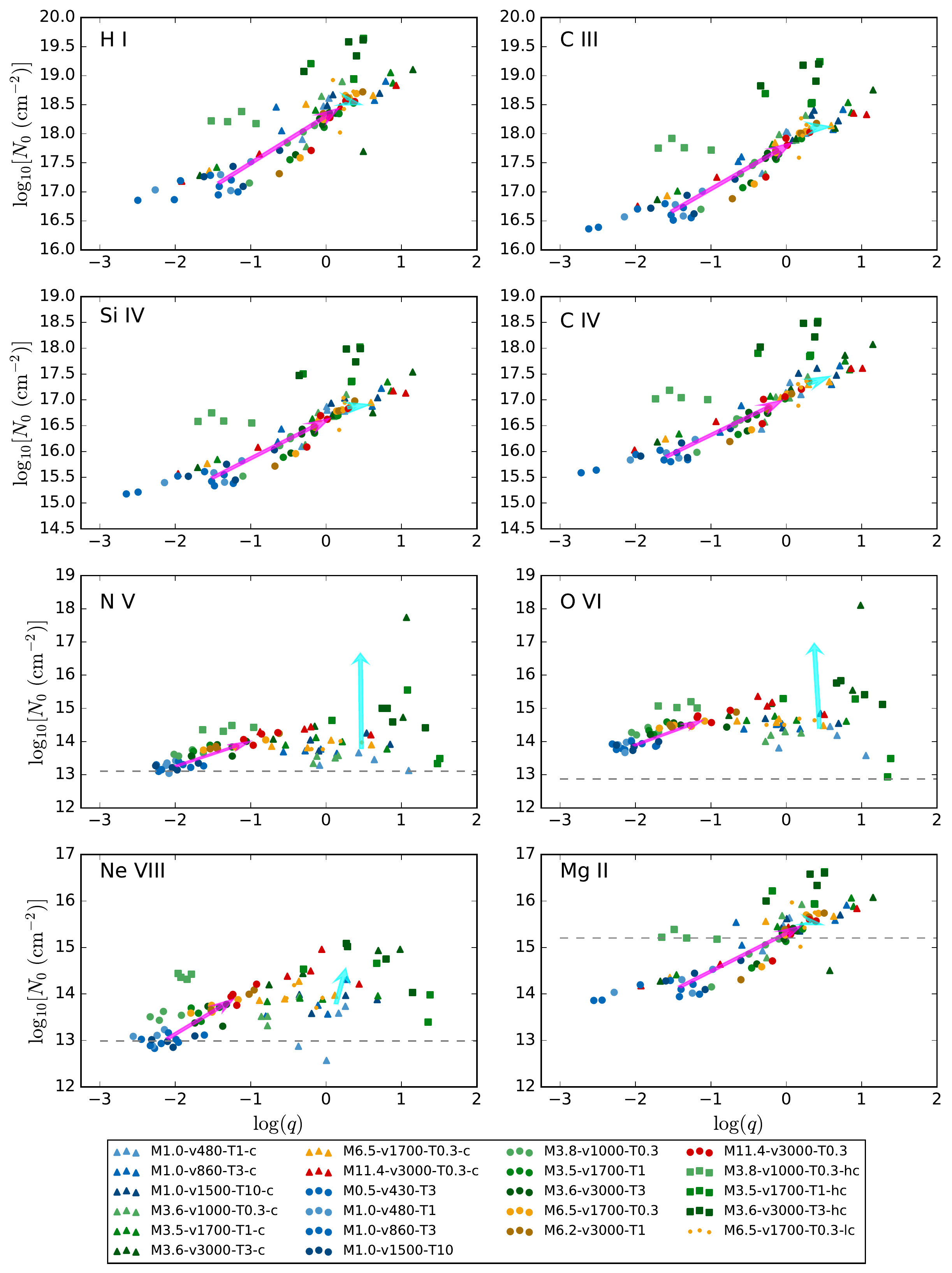}
\caption{Distribution of $N_0$ and $q$ parameters in log space for all runs and 8 ions. Runs are color coded by Mach number with the lowest Mach numbers being in shades of blue and the highest Mach numbers in shades of red. Cooling and conduction runs are marked by circle and triangle points, respectively. Also shown are the high column density runs (squares) and the low conduction run (small dots). Arrows (magenta for cooling and cyan for conduction) are overlaid to highlight the trends discussed in Sections 4.1 and 4.2. Dashed lines indicate observational limits on column densities. \label{fig:jumboMach}}
\end{figure*}

\begin{figure*}
\centering
\includegraphics[angle=0,scale=0.6]{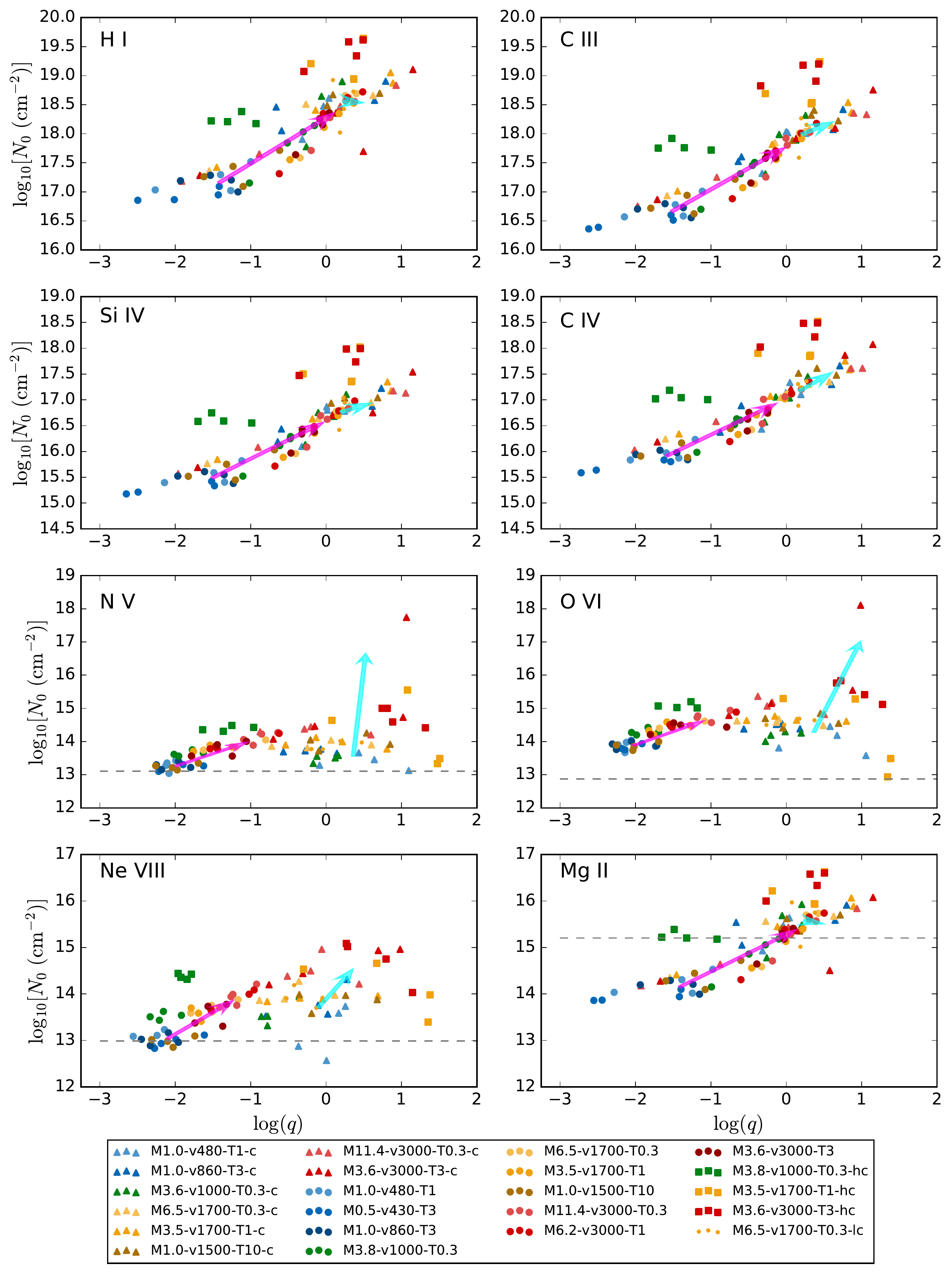}
\caption{Distribution of $N_0$ and $q$ parameters in log space for all runs and 8 ions. Runs are color coded by wind velocity with the lowest velocities in shades of blue and the highest velocities in shades of red. Similar to Figure \ref{fig:jumboMach}, cooling and conduction runs are marked by circle and triangle points, respectively. Also shown are the high column density runs (squares) and the low conduction run (small dots). Arrows (magenta for cooling and cyan for conduction) are overlaid to highlight the trends discussed in Sections 4.1 and 4.2. Dashed lines indicate observational limits on column densities. \label{fig:jumboVel}}
\end{figure*}

The distribution of the best-fit parameters, $N_0$ and $q$, for the down-the-barrel projections are shown in Figures \ref{fig:jumboMach} and \ref{fig:jumboVel}. Of the 10 ions, 8 are shown, excluding \ion{C}{2} and \ion{Si}{3} which are useful for connections to observations but show little to no variation from the distributions seen with \ion{C}{3}. Within both figures, the best-fit parameters for all four times of each run are plotted. The colors are indicative of either the Mach number (Figure \ref{fig:jumboMach}) or wind velocity (Figure \ref{fig:jumboVel}). While these two parameters are related, they affect the results somewhat differently and are the most informative of the simulation parameters. There is no trend with wind temperature. Inefficient conduction runs are denoted with circular markers while runs with conduction are denoted with triangular markers. Trends with increasing Mach number (Fig.\ \ref{fig:jumboMach}) and wind velocity (Fig.\ \ref{fig:jumboVel}) for these subsets are shown with magenta (cooling) and cyan (conduction) arrows. A limiting observable column density can be estimated with the equivalent width, $W = N \lambda f$, where we assume a SNR of 10 and a velocity width of 100 km/s. For a detection at 3 $\sigma$ these column densities range between $\log_{10}(N) \approx 12$ for ions such as \ion{H}{1} and \ion{C}{4} to $\log_{10}(N) \approx15$ for \ion{Mg}{2} and \ion{Si}{3}. They are listed in Table \ref{tab:limitCol}. The best-fit column densities are well above these limits for most ions. However, for the ions where the fits are in the neighborhood of the observation limits, \ion{Mg}{2}, \ion{N}{5}, \ion{O}{6}, and \ion{Ne}{8}, dashed lines have been included in Figures \ref{fig:jumboMach} and \ref{fig:jumboVel} for reference.

\begin{table}
\caption{Estimated limiting column densities for a 3$\sigma$ detection with a SNR of 10 \label{tab:limitCol} }
\centering
\begin{tabular}{cc}
Ion & $\log_{10}N$\\
\hline
\ion{H}{1} & 12.38\\
\ion{Mg}{2} & 15.206\\
\ion{C}{2} & 12.88\\
\ion{Si}{3} & 14.66\\
\ion{Si}{4} & 12.59\\
\ion{C}{3} & 12.11\\
\ion{C}{4} & 12.72\\
\ion{N}{5} & 13.107\\
\ion{O}{6} & 12.87\\
\ion{Ne}{8} & 12.99 
\end{tabular}
\end{table}

\subsection{Conduction vs. Cooling}

Most notably, a majority of the conduction runs span a distinctly different portion of parameter space than the low Mach number runs without conduction. 
In particular, for low ionization ions such as \ion{Mg}{2}, \ion{C}{3}, \ion{C}{4}, and \ion{Si}{4} the runs with inefficient conduction span a lower range of $q$ values than the conduction runs, as demonstrated by the much shorter lengths of cyan arrows for these ions as opposed to the magenta arrows. As low $q$ corresponds to little compaction or high coverage, it is seen clearly here that the cloud material, where much of these ions originate is sparse and diffuse for the runs with inefficient conduction. Additionally for the lower ions, the values for $N_0$ for the inefficient conduction runs tend to be lower than the runs with efficient conduction at low and mid Mach numbers.

For higher ionization ions, \ion{O}{6} and \ion{N}{5}, the distinction between cooling and conduction runs is seen in the $q$ parameter. High values of $q$, corresponding to very compact clouds or little coverage, are dominated by the conduction runs while the runs with inefficient conduction stay within the diffuse cloud regime. If these ions are primarily produced on the boundary of the cloud, as is implied by the fact these ions trace higher temperatures than the core of the cloud (Figure \ref{fig:temphist}), the $q$ parameter for these ions reflects the thickness of these boundaries where these intermediate ions are produced. The cooling and conduction runs have ranges for maximum column density that are consistent with each other. Cooling runs with high velocities appear to have comparable maximum column densities to the conduction runs. It appears that conduction does not significantly influence the amount produced of these ions but may in general produce smaller amounts of coverage.

For the highest ionization energy,\ion{Ne}{8}, there is even more defined clustering between the cooling and conduction runs, primarily dictated by the $q$ parameter. Cooling runs tend toward low $q$ fits while conduction runs exist on the higher end of the $q$ parameter but span a similar range of $N_0$. For runs with inefficient conduction, the \ion{Ne}{8} column density is more dependent on the Mach number while for the conduction runs, the larger column densities correlate with higher velocities.

\subsection{Mach Number and Velocity}

As seen most clearly in Figure \ref{fig:jumboMach}, the Mach number of the wind has a strong influence on the column density profile. Consistently for all ions, the lowest Mach numbers result in the least compact and lowest density profiles and the highest Mach numbers resulting in the most compact clouds. While it is the tendency for high Mach numbers to compress the cloud, which would result in a higher maximum column density, this trend is mostly seen in the runs with inefficient conduction. For the conduction runs, the higher Mach numbers do not influence $N_0$. This is especially apparent when considering that the Mach number trend arrows for conduction do not follow the data as closely as the cooling trend arrows. This is due to conduction runs producing a dense, thin, filament along the flow of the wind which becomes thinner and more extended as the Mach number increases. This is not seen for the runs with inefficient conduction because high Mach numbers produce similar cloudlets to low Mach numbers, but with higher densities.

This compression of the cloud due to higher Mach number winds strongly influences the runs with inefficient conduction. It is most evident in the panels for \ion{N}{5} and \ion{O}{6}. Here there is little change in the amount of each ion produced, as the $N_0$ has little variation. However the best fits for $q$ follow a trend from low to high with increasing Mach number - within the cooling simulations. The compression of the conduction runs is much less dependent on Mach number as the lowest Mach number runs (blue) do not produce notably different best fit parameters than other Mach numbers. The cooling runs that produce the highest $N_0$ and lowest $q$ have moderate Mach numbers, between 3 and 4 (shown in green). This degeneracy is likely do to the fact that the compression for the cooling runs is less significant at higher Mach numbers, allowing for clouds to develop a dense outer layer but an interior with a lower density, ultimately reducing the overall column densities.

The dependence of parameters on velocity is shown in Figure \ref{fig:jumboVel}. Here there is a similar trend to the low Mach numbers, where low velocities (those below 1000 km/s) produce the lowest of the fits for $N_0$ and $q$. The highest velocities result in the maximum $N_0$ for both cooling and conduction and the trend arrows for both efficient and inefficient conduction appear to follow the general shape of the data. High velocities also correspond to high $q$ values, reflecting the effects of shocks on both the cloudlets within a cooling run and the filaments in the conduction runs to compress the gas.

\section{Application: Down-the-Barrel Outflow Observations}\label{sec:app}

As an illustration of the types of studies enabled by our results, we consider an application of our tabulated fits. We consider the observations in \citet{Chisholm2017} in particular as many of our assumptions, including radiation background and ionization parameter, are derived from these observations. \citet{Chisholm2017} report absorption profiles from down-the-barrel observations, of a lensed galaxy with $z \approx 2.9$. Observations are made of both low and high ionization profiles that indicate the two phases are co-spatial, much like the wind-cloud interaction considered here. We aim to determine an appropriate scaling of our column density profiles by accounting for two factors that can influence the optical depth of the absorbing clouds. 

To approximate an absorption profile for these ions we estimate the observed optical depth for a particular ion as
\begin{equation}
\tau(v) = \frac{c \sigma}{\sqrt{\pi} b} \exp \left[-\frac{(v-v_0)^2}{b^2} \right] \, \bar N, \label{eq:tau}
\end{equation}
where $\bar N \equiv \frac{\sum_i^x N_i}{x},$ is the average number density of the ion, where $x$ is the number of points within the column density profile and $N_i$ the column density for the $i$th point. Here $\tau$ is approximated with the center of the absorption profile at the average velocity of the cloud, $v_0$, with a velocity dispersion determined by the $b$ parameter estimated for each ion and $v$ is the velocity bin within the absorption profile. We consider velocities between -600 km/s to 200 km/s offset from line center. This average optical depth approximates a single cloud.

There are then two ways to parameterize the absorption profile, we consider the two parameters independently. The first method is the altering covering fraction which describes the fraction of the area within the sightline that is obscured by the cloud. With this covering fraction parameter, the observed flux from the derived column density profiles can be expressed as
\begin{equation}
F (v) = (1-f) + f e^{-\tau(v)},
\end{equation}
where $\tau(v)$ is the average optical depth above and $f$ is the free parameter describing the covering fraction. 

The second way to parameterize the absorption profile considers scaling the optical depth either to represent multiple clouds along the sightline or one cloud with scaled density. For this we assume the intervening cloud has the average optical depth of a single cloud from Equation \ref{eq:tau}, which is then scaled by the parameter $\alpha$ which describes the number of clouds within the sightline or the scaling factor of the density for a single cloud. In this case, the observed flux can be described as
\begin{equation}
F (v) = e^{-\alpha \tau(v)},
\end{equation}
where $\tau(v)$ is again the average optical depth above and $\alpha$ is the second free parameter we consider, describing the number of clouds.

Best-fit covering fraction and $\alpha$ are found by performing $\chi^2$ minimizations for each of the parameters independently over the five ions shared between our analysis of the \citet{Chisholm2017} observations, \ion{C}{2}, \ion{C}{4}, \ion{Si}{4}, \ion{O}{6} and \ion{N}{5}. We determine this $\chi^2$ for each of the four evolutionary stages, which here correspond to a central velocity, for all of the 22 runs. As each of these stages could be observed in a single observation when looking down the barrel of an outflow, we consider the average and maximum $\chi^2$ between these stages to determine the goodness of fit for each simulation.

\begin{figure}
\includegraphics[angle=0,width=\columnwidth]{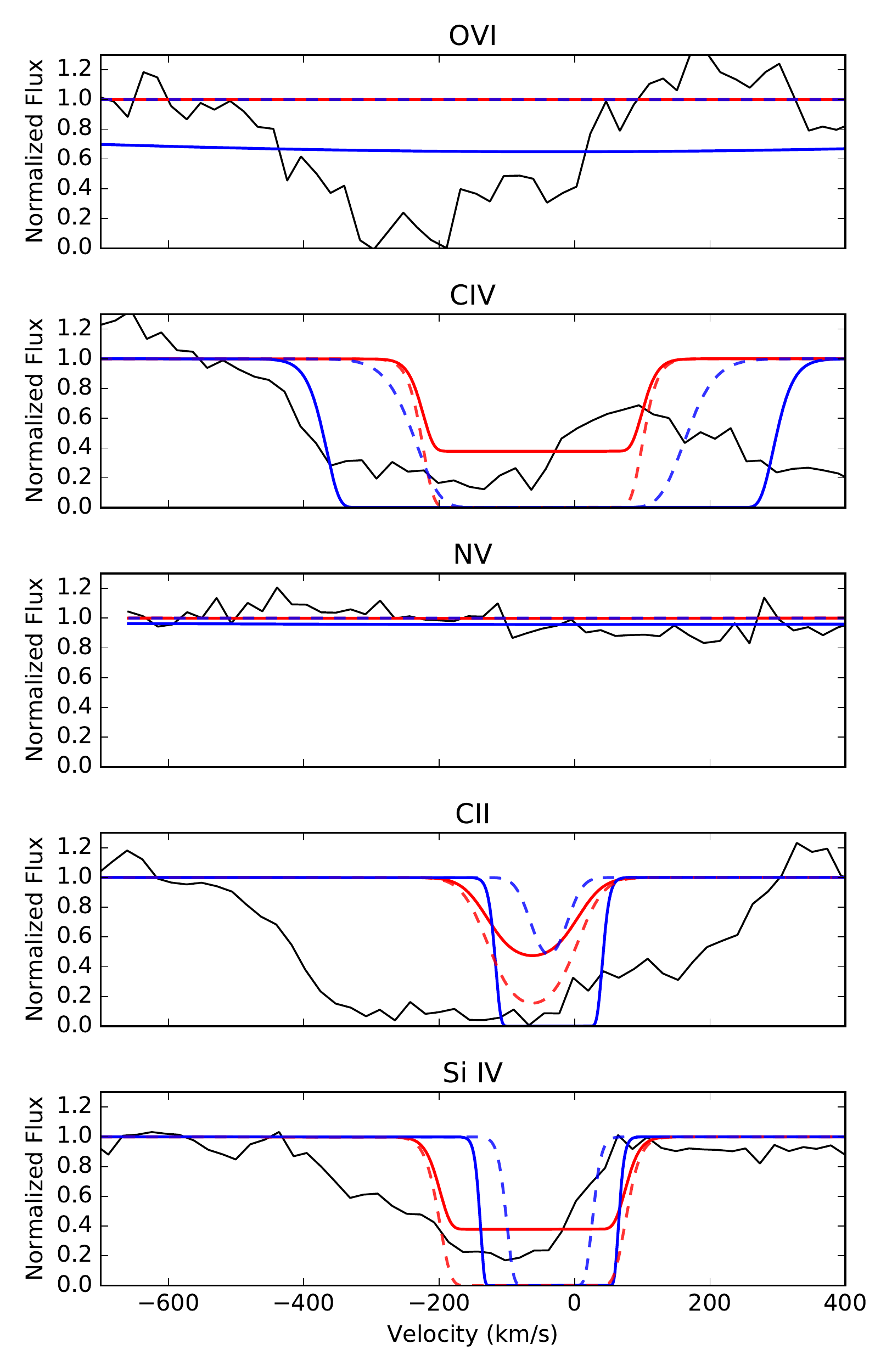}
\caption{The best fit absorption profiles to the \citet{Chisholm2017} observations (shown in black) for the multi-cloud (blue) and covering fraction (red) cases shown for one characteristic time, $t_{25}$. The covering fraction best-fits correspond to the run M3.5-v1700-T1. The multi-cloud best-fits correspond to the run M3.6-v3000-T3-hc. Also shown in dashed lines are the profiles with $\alpha = 1$ (blue dashed) and $f = 1$ (red dashed) for the same runs. \label{fig:outflowFit}}
\end{figure}

\begin{table*}
\caption{Best fit $N_0$ and $q$ for high resolution run at each stage \label{tab:highres} }
\centering
\begin{tabular}{ccc}
& M3.5-v1700-T1 & M3.5-v1700-T1-highres\\
\hline
$q_1$ & -1.756 & -1.804\\
$q_2$ & -1.820 & -2.097\\
$q_3$ & -1.583 & -1.848\\
$q_4$ & -1.272 & --\\
\hline
$\log (N_{0, 1})$ & 14.208 & 14.040\\
$\log (N_{0, 2})$ & 14.328 & 14.165\\
$\log (N_{0, 3})$ & 14.588 & 14.275\\
$\log (N_{0, 4})$ & 14.490 & --
\end{tabular}
\end{table*}

The best fit for both approaches, each with their own free parameter, is shown in Figure \ref{fig:outflowFit}, with the covering fraction fits in red and the multi-cloud fits in blue. Also shown with dashed lines is the $\alpha = 1$ and $f=1$ cases, highlighting the profiles produced with the base case of one cloud with full coverage over the sightline. The run with the best-fit for the covering fraction case is M3.5-v1700-T1-c with a covering fraction of 0.621. The covering fraction approach is able to generate profiles that approximate the nearly saturated lines \ion{Si}{4} and \ion{C}{4}, and maintain the low levels of \ion{N}{5} to match observations in \citet{Chisholm2017}. However, there is not enough \ion{C}{2} or \ion{O}{6} in our simulations to reproduce the observed profiles with a single cloud.

In case in which the optical depth is increased uniformly, parameterized by the number of clouds, we find a best-fit $\alpha$ of 488 for the best-fit run M3.6-v3000-T3-hc. It is important to acknowledge that it is highly unlikely that 488 clouds would be lined up to each fully cover a particular sightline, though scaling the density of one cloud by this factor is feasible. However this number, paired with the fact that this best-fit run is one of the high column density runs, demonstrates that there is a significant discrepancy between these simulations and the observations. In particular, our simulations do not produce enough \ion{C}{2}. While \ion{Si}{4} and \ion{C}{4} are saturated in the base case and more clouds only widen the profile, the low levels of \ion{C}{2} drive up the number of clouds necessary in order to approach the nearly saturated \ion{C}{2} observations.

For the intermediate ions, \ion{Si}{4}, \ion{C}{4} and \ion{C}{2} it is also clear that the derived line widths are much narrower than those observed in \citet{Chisholm2017}. This limits the simulations' potential to produce these wide profiles by simply altering the optical depth. However for \ion{O}{6}, there is an opposite effect. While most of the high column density \ion{O}{6} is found on the interface of the cloud, there is a portion of \ion{O}{6} that can be found in the hot wind (see Figures \ref{fig:projections} and \ref{fig:temphist}). This gives \ion{O}{6} velocity dispersions on the order of the wind velocity, $10^3$ km/s. This can account for the nearly flat appearance of the best-fit for the multiple cloud approach in \ion{O}{6}. The shallow and wide profile appears flat over the range of velocities relevant for the other ions. Even so, there is not enough \ion{O}{6} within or on the interfaces of the slower moving cloud to reproduce the deep profiles observed. It is possible that this discrepancy could be explained by the effects of low resolution. However Table \ref{tab:highres} shows the best-fit parameters for \ion{O}{6} a higher resolution cooling run which would be most affected by resolution effects due to the fact \ion{O}{6} is produced within mixing layers. The higher resolution run produces comparable maximum column densities and similar coverage parameters, $q$, to the run with the resolution used throughout the rest of the simulations. 

Ultimately, a more realistic view would treat both of these factors together, introducing the influence of a density scaling or multiple clouds each with their own covering fraction. However, these simplistic views can support the need for further investigation. While comparable amounts of \ion{Si}{4} and \ion{C}{4} absorption can be recreated, the lack of \ion{C}{2} indicates there is a significant factor not accounted for that enables more cold cloud material to remain within the sightline throughout the interaction with the wind. The wide velocity dispersions of \ion{O}{6} also indicate a need to determine a source of \ion{O}{6} ionization that can introduce noticeable absorption over a smaller velocity range.

\section{Discussion and Summary} \label{sec:discuss}

Starburst-driven galactic outflows are a complex, multiphase phenomenon, and understanding their evolution requires close comparisons between observations and numerical studies. While numerical simulations can reproduce the full evolution of cold clouds interacting with hot wind material given a set of assumptions about the underlying physical processes, observations are often limited to absorption line profiles of species with low and intermediate ionization states.

To help in interpreting such observations, we have derived equilibrium column density profiles, average velocities, and $b$ parameters for 22 hydrodynamical simulations of cold, atomic clouds in super-sonic winds including both radiative cooling and thermal conduction. These capture the equilibrium distributions of ten widely-observed ions: \ion{H}{1}, \ion{Mg}{2}, \ion{C}{2}, \ion{C}{3}, \ion{C}{4}, \ion{Si}{3}, \ion{Si}{4}, \ion{N}{5}, \ion{O}{6}, and \ion{Ne}{8}. With the possible exception of \ion{H}{1}, the column density profiles are all well fit by the functional form $N(x) = N_0 \frac{0.01}{1.01-x^q},$ where $x$ is the fractional rank of each cell expressed as a fraction of the total, $N_0$ places a upper limit on column density fand $q$ expresses the degree to which the cloud has been compacted. For all ions we provide tabulated fits of $N_0$ and $q$ for each simulation case, at four characteristic times.

As a general trend, the runs including conduction produce much higher column densities and much narrower amounts of coverage, coinciding with the more compact, dense filaments produced in late stages of the cloud-wind interaction. The runs with inefficient conduction are restricted to lower column densities for most ions except \ion{N}{5} and \ion{O}{6}, which are primarily produced at the cloud-wind boundary. These runs also follow more predictable trends with functions of wind velocity and Mach number as higher velocities compact the cloud and result in higher column densities.

Our study is limited by the need to reduce the parameter space with the assumption that the metallicity is solar. While this is an estimate of the maximum metallicity within the CGM, the absorption observed near starbursts is more likely due to high amounts ion ionization rather than a high metal content. We also assume a radiation model representative of a young starburst galaxy with at a high ionization parameter ($\log U = -2$), which can greatly vary between CGM observations and is a necessary component to making connections to CGM observations such as COS-Burst. 

However, even in comparison to observations that best match our assumed parameters, we find that we cannot reproduce observed absorption line column density ratios with our equilibrium model. Our results overestimate the amounts of intermediate ions such as \ion{Si}{4} and \ion{C}{4}, as they produce saturated profiles. Due to this, the best fit parameters that produce fits that closely match the profiles for \ion{Si}{4} and \ion{C}{4} also significantly underestimate the absorption from \ion{O}{6}, \ion{N}{5} and \ion{C}{2}. The discrepancy between \ion{O}{6} and \ion{N}{5} absorption is also not explained by the inclusion of thermal conduction. In both cases, inefficient and efficient conduction, the column densities of both ions are comparable and not impacted by resolution effects. Thus it is possible that the lack of \ion{N}{5} observed is linked to non-equilibrium processes, \citep[e.g.][]{Grassi2014,Gray2015,Gray2016,Gray2017, Pallottini2017}, which must be accounted for through the use of a full chemical network.

Addressing this issue will require performing a similar analysis on outflow simulations including non-equilibrium chemistry. These can then be compared with the present fits to demonstrate the drawbacks of equilibrium assumption, and they will yield better estimates of the abundances of each ion. Other consideration should be given to the effects of different ionization parameters and metallicities that are more descriptive of the CGM, as well as the balance between cooling and potential photo-heating. These parameters are likely to have a significant effect on the production of low ions in particular. Similarly, simulations including other effects such as the impact of magnetic fields and cosmic rays \citep[e.g.][]{Simpson2016,Ruszkowski2017,Fujita2018,Samui2018}, as well as addressing the contribution of cold gas created in the expanding wind \citep[e.g.][]{Thompson2016, Scannapieco2017, Schneider2018} will likely be needed to fully address the parameter space of physical process impacting galactic outflows, their interaction with the CGM, and their influence on galaxy evolution.

\acknowledgments
This work was supported by the National Science Foundation under grant AST14-07835 and NASA theory grant NNX15AK82G. We would like to thank John Chisholm, Sanchayeeta Borthakur and Neal Katz for informative and helpful discussions as well as Cameron Hummels and Britton Smith for advice regarding the use of \textsc{trident}. ES gratefully acknowledges the Simons Foundation for funding the workshop Galactic Winds: Beyond Phenomenology which helped to inspire this work. We would also like to thank the Texas Advanced Computing Center (TACC) at The University of Texas at Austin, and the Extreme Science and Engineering Discovery Environment (XSEDE) for providing HPC resources via grant TGAST130021 that have contributed to the results reported within this paper. MB is supported by a grant by the Deutsche Forschungsgemeinschaft under BR2026125.

\bibliography{mybib}

\begin{table*}
\caption{Sample of table of best fit parameters $q$ and $\tau$ -- \ion{C}{4} \label{tab:paramSample}}
\centering
\begin{tabular}{cllllllllll}
Run & Velocity & b &$N_0$ & Upper & Lower & q & Upper & Lower & Average N & Average N\\
& (km/s) & (km/s) & (cm$^{-2}$) & $N_0$ Err & $N_0$ Err & Fit & q Err & q Err &(cm$^{-2}$) & Err\\
\tableline
\tableline
M3.8-v1000-T0.3-c & 22 & 971 & 2.813e+17 & 3.179e+9 & 2.813e+17 & 1.925 & 24.641 & 4.180e-08 & 8.586e+15 & 7.410e+6\\
M3.8-v1000-T0.3-c & 36 & 912 & 1.140e+17 & 1.424e+9 & 1.140e+17 & 0.725 & 20.688 & 1.559e-08 & 6.507e+15 & 2.321e+7\\
M3.8-v1000-T0.3-c & 61 & 869 & 1.091e+17 & 9.405e+8 & 1.091e+17 & 1.099 & 19.531 & 1.628e-08 & 4.734e+15 & 2.467e+7\\
M3.8-v1000-T0.3-c & 110 & 777 & 3.724e+16 & 3.396e+8 & 3.724e+16 & 0.517 & 28.297 & 8.347e-09 & 2.666e+15 & 2.821e+6\\
M3.6-v3000-T3-c & 21 & 62 & 1.535e+16 & 1.260e+13 & 1.535e+16 & 0.019 & 47.396 & 6.820e-07 & 7.185e+15 & 3.632e+6\\
M3.6-v3000-T3-c & 14 & 36 & 7.317e+17 & 8.702e+10 & 7.317e+17 & 6.007 & 20.083 & 1.640e-3 & 1.252e+16 & 3.824e+7\\
M3.6-v3000-T3-c & 43 & 80 & 1.610e+17 & 2.122e+9 & 1.610e+17 & 1.155 & 26.605 & 2.650e-08 & 6.765e+15 & 4.003e+7\\
M3.6-v3000-T3-c & 39 & 73 & 1.185e+18 & 2.859e+10 & 1.185e+18 & 14.060 & 14.276 & 1.136e-1 & 1.549e+16 & 1.840e+8\\
M3.5-v1700-T1-c & 13 & 118 & 2.196e+16 & 4.838e+11 & 2.196e+16 & 0.037 & 49.787 & 1.229e-05 & 7.629e+15 & 2.745e+6\\
M3.5-v1700-T1-c & 21 & 116 & 1.182e+17 & 1.154e+9 & 1.182e+17 & 0.613 & 23.710 & 7.903e-09 & 7.548e+15 & 7.335e+7\\
M3.5-v1700-T1-c & 27 & 143 & 5.673e+17 & 5.985e+9 & 5.673e+17 & 6.032 & 13.455 & 2.156e-06 & 9.695e+15 & 3.111e+7\\
M3.5-v1700-T1-c & 36 & 107 & 3.829e+17 & 5.077e+9 & 3.829e+17 & 6.894 & 11.566 & 2.551e-3 & 6.215e+15 & 6.473.e+3\\
M3.8-v1000-T0.3 & 108 & 787 & 4.291e+16 & 5.257e+9 & 4.291e+16 & 0.220 & 36.217 & 1.331e-08 & 5.399e+15 & 2.070e+7\\
M3.8-v1000-T0.3 & 143 & 679 & 4.051e+16 & 6.549e+9 & 4.051e+16 & 0.245 & 26.992 & 2.333e-09 & 4.755e+15 & 1.464e+7\\
M3.8-v1000-T0.3 & 158 & 637 & 3.315e+16 & 3.066e+8 & 3.315e+16 & 0.205 & 32.308 & 2.917e-08 & 4.352e+15 & 1.893e+7\\
M3.8-v1000-T0.3 & 204 & 613 & 9.625e+15 & 3.212e+13 & 9.625e+15 & 0.065 & 47.141 & 8.882e-06 & 2.514e+15 & 2.935e+7\\
\vdots & \vdots & \vdots & \vdots & \vdots & \vdots & \vdots & \vdots & \vdots &\vdots & \vdots
\end{tabular}
\end{table*}

\clearpage

\end{document}